\documentclass[%
 reprint,
showpacs,preprintnumbers,
 amsmath,amssymb,
 aps,
]{revtex4-1}
\usepackage{graphicx}
\usepackage{color}
\usepackage{amsmath}
\usepackage{dcolumn}
\usepackage{bm}

\begin{document}

\preprint{APS/123-QED}

\title{Electron and phonon transport in shandite-structured Ni$_3$Sn$_2$S$_2$}

\author{ Alex Aziz, Panagiotis Mangelis, Paz Vaqueiro, Anthony V. Powell and Ricardo Grau-Crespo*}

\affiliation{Department of Chemistry, University of Reading, Whiteknights, Reading RG6 6AD, United Kingdom.}
\email{r.grau-crespo@reading.ac.uk}
\date{\today}

\begin{abstract}
The shandite family of solids, with hexagonal structure and composition \textit{A}$_3$\textit{M}$_2$\textit{X}$_2$ (\textit{A} = Ni, Co, Rh, Pd, \textit{M} = Pb, In, Sn, Tl, \textit{X} = S, Se), has attracted recent research attention due to promising applications as thermoelectric materials. Herein we discuss the electron and phonon transport properties of shandite-structured Ni$_3$Sn$_2$S$_2$, based on a combination of density functional theory (DFT), Boltzmann transport theory, and experimental measurements. Ni$_3$Sn$_2$S$_2$ exhibits a metallic and non-magnetic groundstate with Ni$^0$ oxidation state and very low charge on Sn and S atoms. Seebeck coefficients obtained from theoretical calculations are in excellent agreement with those measured experimentally between 100 and 600 K. From the calculation of the ratio $\sigma$/$\tau$ between the electronic conductivity and relaxation time, and the experimental determination of electron conductivity, we extract the variation of the scattering rate (1/$\tau$) with temperature between 300 and 600 K, which turns out to be almost linear, thus implying that the dominant electron scattering mechanism in this temperature range is via phonons. The electronic thermal conductivity, which deviates only slightly from the Wiedemann-Franz law, provides the main contribution to thermal transport. The small lattice contribution to the thermal conductivity is calculated from the phonon structure and third-order force constants, and is only $\sim$2 Wm$^{-1}$K$^{-1}$ at 300 K (less than 10\% of the total thermal conductivity), which is confirmed by experimental measurements. Overall, Ni$_3$Sn$_2$S$_2$ is a poor thermoelectric material ($ZT\sim$ 0.01 at 300 K), principally due to the low absolute value of the Seebeck coefficient. However, the understanding of its transport properties will be useful for the rationalization of the thermoelectric behavior of other, more promising members of the shandite family.
\end{abstract}


\maketitle

\section{\label{sec:level1}Introduction\protect\\}

Chalcogenides of general formula \textit{A}$_3$\textit{M}$_2$\textit{X}$_2$ (A = Ni, Co, Rh, Pd;  M = Pb, In, Sn, Tl, Bi;  X = S, Se) exhibit interesting electronic and magnetic properties including  superconductivity (Ni$_3$Bi$_2$S$_2$) \cite{sakamoto_06}, half-metallic ferromagnetism (Co$_3$Sn$_2$S$_2$) \cite{Weihrich_04, Ferro_Vaqueiro}, and metal-insulator transitions (Co$_3$Sn$_{2-x}$In$_x$S$_2$) \cite{Weihrich_06}. The latter series has also been recently investigated for its potential for thermoelectric applications at high temperature ~\cite{corps_2013,corps_interplay}. In particular, the thermoelectric figures of merit reported for this solid solution at 425 K, when 0.8 $< x <$ 0.85,  is among the highest reported for sulfide phases in this temperature range, suggesting it may have applications in low-grade waste heat recovery. 

The shandite structure, adopted by the Co$_3$Sn$_{2-x}$In$_x$S$_2$ series at all compositions (as well as by most compounds with \textit{A}$_3$\textit{M}$_2$\textit{X}$_2$ stoichiometry), consists of sheets of metal atoms (both \textit{A} and \textit{M}) in the form of a Kagome-like hexagonal network,  capped above and below by \textit{X} atoms, and  stacked in \textit{ABC} sequence. There is a second \textit{M} site, located between the Kagome sheets, with trigonal anti-prismatic coordination to the \textit{X} atoms (Fig. 1). The distribution of Sn and In over the two types of $M$ sites has been found to be an important factor in the explanation of the electronic behavior of the Co$_3$Sn$_{2-x}$In$_x$S$_2$ solid solution \cite{corps_interplay}. 

The present study focuses on understanding the electron and phonon transport properties of Ni$_3$Sn$_2$S$_2$, as a representative of the shandite family. Contrasting with the magnetic nature of Co shandites, Ni$_3$Sn$_2$S$_2$ has a non-magnetic ground state, which has been confirmed by band structure calculations and photoelectron spectroscopy \cite{gutlich_valence}, as well as by direct magnetic susceptibility measurements \cite{kubodera}. The presence of spin polarization and magnetic excitations complicates the calculation of transport coefficients, and also the theoretical description of electron scattering, as electron-magnon interactions have to be taken into account \cite{spin, Graf09}. The absence of magnetism in Ni$_3$Sn$_2$S$_2$ thus makes this compound a convenient starting point for a theoretical investigation of transport phenomena and thermoelectric behavior in shandites. 

In addition to the results of electronic and phonon structure calculations, we present here theoretical predictions as well as experimental measurements of all the transport coefficients contributing to the thermoelectric figure of merit:
\begin{equation}
    ZT=\frac{\sigma S^2 T}{\kappa_{\text{el}}+ \kappa_{\text{latt}}}\\
\end{equation}

i.e, the Seebeck coefficient $S$, the electrical conductivity $\sigma$, and the electronic ($\kappa_{\text{el}}$) and lattice ($\kappa_{\text{latt}}$) contributions to the thermal conductivity. We will examine the variation with temperature ($T$) of each of the coefficients and discuss the physical mechanisms responsible for the transport behavior.

\begin{figure}[!ht]
\centering
\includegraphics[width=90mm]{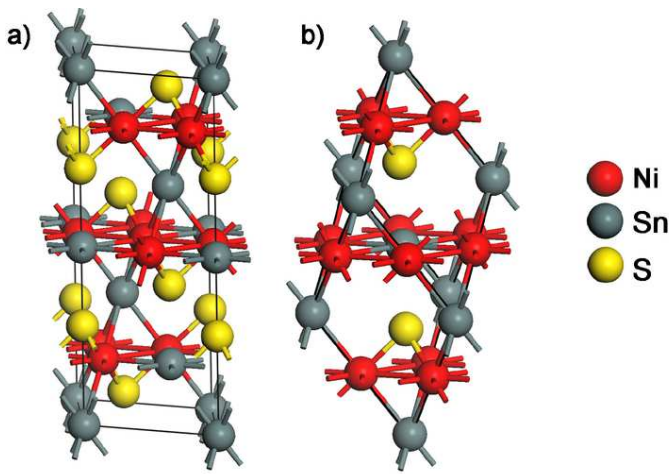}
\caption{Hexagonal (a) and rhombohedral (b) unit cell of shandite-structured Ni$_3$Sn$_2$S$_2$.}
\end{figure}

\section{Methodology}

\subsection{Computational techniques}

\subsubsection{Density functional theory calculations}
The crystal structure of Ni$_3$Sn$_2$S$_2$ was optimised using periodic density functional theory (DFT) calculations as implemented in the Vienna Ab initio Simulation Package (VASP)~\cite{kresse_96a,kresse_96b}. The projector augmented wave (PAW) method was used \cite{bloch94, kresse99paw}, with electron levels up to Ni 3\textit{p}, Sn 4\textit{p} and S 2\textit{p} kept frozen at their reference atomic state. The exchange-correlation functional of Perdew-Burke-Ernzerhof (PBE) \cite{perdew_pbe}, based on the generalized gradient approximation (GGA), was employed. The number of plane waves was determined using a kinetic energy cutoff of 350 eV. Reciprocal space integrations were performed on a $\Gamma$-centered grid of k-points with the smallest permitted spacing between them of 0.3 $\text{\AA}^{-1}$, which corresponds to a $7\times7\times7$ grid on the reciprocal lattice of the primitive cell.  Spin-polarized calculations were performed with different initializations of the magnetic moments but the calculations always converged to a non-magnetic groundstate, as expected from previous research \cite{gutlich_valence}.  The ionic positions were relaxed until the forces were less than 0.01 eV$\text{\AA}^{-1}$ on each atom.  A Bader analysis \cite{bader_atoms} of the charge density from VASP was performed using the code by Henkelman et al. \cite{henkelman_06, tang_2009}.

\subsubsection{Electron transport calculations}
As a starting point for the electronic transport calculations, we re-determined the band structure using the WIEN2k code \cite{wien2k_blaha}. For this calculation, the Brillouin zone was sampled with a fine k-mesh of 50$\times$50$\times$50 points. For the basis set expansions we used the cutoff parameters $l_{\text{max}}=10$ and $R_{\text{mt}}K_{\text{max}}=7$, while for the charge density Fourier expansion we used the cutoff $G_{\text{max}}=12$; all these parameters were checked for convergence of the total energy. The radii of the muffin-tin spheres were set at the default values of 2.26, 2.50 and 1.85 bohrs for Ni, Sn, and S, respectively. The transport coefficients were then obtained from the bands by solving the linearized Boltzmann transport equation using the BoltZTraP code \cite{madsen_boltz}, which interfaces with the WIEN2k output. BoltZTraP uses the relaxation time approximation and a ``rigid band'' approach to obtain the transport coefficients as functions of the electron chemical potential and temperature. Both the electrical conductivity ($\sigma$) and the electronic contribution to the thermal conductivity are calculated relative to the relaxation time ($\tau$), which is assumed to be isotropic and constant in the reciprocal space at each temperature. The Seebeck coefficient can be calculated on an absolute scale, i.e. it is independent of $\tau$. The temperature variation of $\tau$ is discussed based on the comparison with experimental measurements of the electrical conductivity.  At each temperature, we use the equilibrium value of the chemical potential corresponding to the undoped system, which deviates only slightly from the Fermi level. However, we also consider the effects of (dilute) doping by evaluating the transport coefficients and $ZT$ at different chemical potentials corresponding to different concentrations of electron/hole doping.   

\subsubsection{Phonon transport calculations}
In order to calculate the lattice contribution $\kappa_{\text{latt}}$ to the thermal conductivity, we solved the phonon Boltzmann transport equation using the method implemented in the ShengBTE code \cite{shengbte_1,thermal_nanowire_2012}, which goes beyond the relaxation time approximation (RTA) to provide a full iterative solution. The method requires the calculation of both second-order (harmonic) and third-order (anharmonic) force constants, which were obtained by the finite-displacement method, using energies from VASP calculations in a $3\times3\times3$ supercell of the primitive cell. The phonon dispersion curves and heat capacity were obtained from the second-order force constants using the Phonopy code \cite{togo_2008}. For the efficient calculation of the anharmonic force constants, harnessing the crystal symmetry, we use the thirdorder.py script \cite{thermal_nanowire_86}. This required the evaluation of the DFT energies of 364 configurations of atom displacements. The ShengBTE calculations were performed using a q-point grid of $13\times13\times13$, which was tested for convergence. 

\subsection{Experimental techniques}

\subsubsection{Sample preparation}
 Ni$_3$Sn$_2$S$_2$ was synthesized by the sealed tube method at high temperatures. Mixtures of elemental nickel (Alfa, powder, 99.9\%), tin (Aldrich, powder, ≥ 99\%), and sulphur (flakes, Aldrich, 99.99\%) were ground using an agate pestle and mortar. The resulting powders were sealed under vacuum (between $10^{-3}$-$10^{-4}$ mbar) into a fused silica tube, and the mixture fired for two periods of 48 h at $500\,^{\circ}\mathrm{C}$ and $700\,^{\circ}\mathrm{C}$ with an intermediate regrinding. A heating and cooling rate of $0.5\,^{\circ}\mathrm{C}$ $\text{min}^{-1}$ was used.

\subsubsection{Powder X-ray Diffraction}
The structural characterization was carried out by powder X-ray diffraction (XRD) using a Bruker D8 Advance diffractometer (Ge-monochromated Cu K$\alpha_1$, $\lambda$ = 1.5406 $\text{\AA}$ and a LynxEye linear detector. Rietveld refinement of lattice, atomic, thermal and profile parameters was carried out using the GSAS software package \cite{gsas}. 

\subsubsection{Measurement of transport properties}
The densification of the powdered sample was carried out using a hot press manufactured in-house. After ball milling at 350 rpm for 1 hour, a sample with a mass of about 1.8 g was loaded between two graphite dies in a graphite mold. Hot-pressing under a N$_2$ atmosphere, at 60 bar and 995 K for 25 min, led to a pellet with a density of 99.6\% relative to the bulk material. The density of the resulting pellets was measured by the buoyancy method. Thermal diffusivity measurements at high temperatures 300 $\leq$ T(K) $\leq$ 525 were conducted using an LFA 447 Nanoflash Netzsch instrument, and the determination of the specific heat carried out via the comparison method. Pyroceram 9606 was used as the reference sample. The electrical resistivity and Seebeck coefficient were measured at high temperatures (300 - 670 K) using a Linseis LSR3-800 instrument. Low-temperature (100 - 300 K) Seebeck coefficient measurements were conducted in 10 K steps using an in-house instrument equipped with a close-cycle refrigerator.

\section{\label{sec:level1}Results and discussion\\}

\subsubsection{Crystal structure}
The XRD analysis confirms that Ni$_3$Sn$_2$S$_2$ crystallizes in the rhombohedral {\it R\=3m} space group, as reported by Range et al \cite{cif}.
The lattice parameters obtained from the DFT optimization were in good agreement with experiment, and with literature values \cite{cif} (Table I). The small overestimation of the lattice parameters is typical of GGA-PBE calculations of metallic systems \cite{haas}. Another source of discrepancy is that the DFT results are obtained by minimization of the total energy at zero temperature (or more precisely, without considering vibrational effects, as zero-point effects were not included either), while the reported experimental parameters were measured near room temperature (293 K in this work and 297 K in Ref. \cite{cif}). Still, the discrepancies are very small (+1.4\% for $a$ and +0.15\% for $c$).

\begin{table}[b]%
\caption{\label{tab:table1}%
Comparison of theoretical and experimentally determined crystal parameters of Ni$_3$Sn$_2$S$_2$. $z$[S] is the $z$ fractional coordinate of the S atom in special position 6c (0, 0, $z$) of space group {\it R\=3m}.
}
\begin{ruledtabular}
\begin{tabular}{lcccc}
\multicolumn{1}{c}{ } &
 \multicolumn{2}{c}{This work} &
 \multicolumn{1}{c}{Ref \cite{cif}} \\
\textrm{Parameter}&
\textrm{DFT}&
\multicolumn{1}{c}{\textrm{Experiment}}&
\multicolumn{1}{c}{\textrm{Experiment }}\\
\colrule
\textit{a} ($\text{\AA}$) & 5.540 & 5.46771(7) & 5.4606(2) \\
\textit{c} ($\text{\AA}$) & 13.208 & 13.1922(2) & 13.188(1)\\
\textit{V} ($\text{\AA}^{3}$)& 351.06 & 341.55 & 340.56 \\
\textit{z}[S] & 0.2792 & 0.2820(2) & 0.2820(2) \\
\end{tabular}
\end{ruledtabular}
\end{table}

 \subsubsection{Electronic structure}
The electronic band structure between high-symmetry points is shown in Fig. 2.  Ni$_3$Sn$_2$S$_2$ has a non-magnetic metallic groundstate with a low density of states at the Fermi level. The projection of the density of states on the Ni  3\textit{d}-orbitals shows that these contributions are almost completely below the Fermi level, which indicates a Ni$^0$ formal oxidation state with a 3\textit{d}$^{10}$ configuration. The 4\textit{s} orbitals are about 5 eV above the Fermi level. The 3\textit{d}$^{10}$ configuration is typical of Ni$^0$ in inorganic molecular compounds like Ni(CO)$_4$  \cite{bauschlicher_Ni0,mckinlay_Ni0}. The neutral state of Ni is consistent with the Bader analysis which is shown in Table II. The  Ni$^0$ valence state and the nature of the groundstate are also in agreement with the findings in Ref. \cite{gutlich_valence}. 

The magnitudes of the charges associated with the Sn and S atoms ($\pm$0.6-0.7) are also well below what would be expected from formal oxidation states, but still significantly different from zero: they are in between those found for SnS (polar covalent compound) and SnSb (intermetallic compound). 

It is interesting to compare the electronic structure of Ni$_3$Sn$_2$S$_2$ with that of Co$_3$Sn$_2$S$_2$. The latter has a ferromagnetic groundstate with half-metallic character, exhibiting a gap of ca. 0.3 eV for the minority spin channel \cite{schnelle}. Despite that fundamental difference, the total charge density distribution over the atoms is very similar for both compounds. In Co$_3$Sn$_2$S$_2$, Co is found to be zero-valent, while the atomic charges for Sn/S are also very close to those found for Ni$_3$Sn$_2$S$_2$ \cite{corps_interplay}.

\begin{figure}[b]
\centering
\includegraphics[width=90mm]{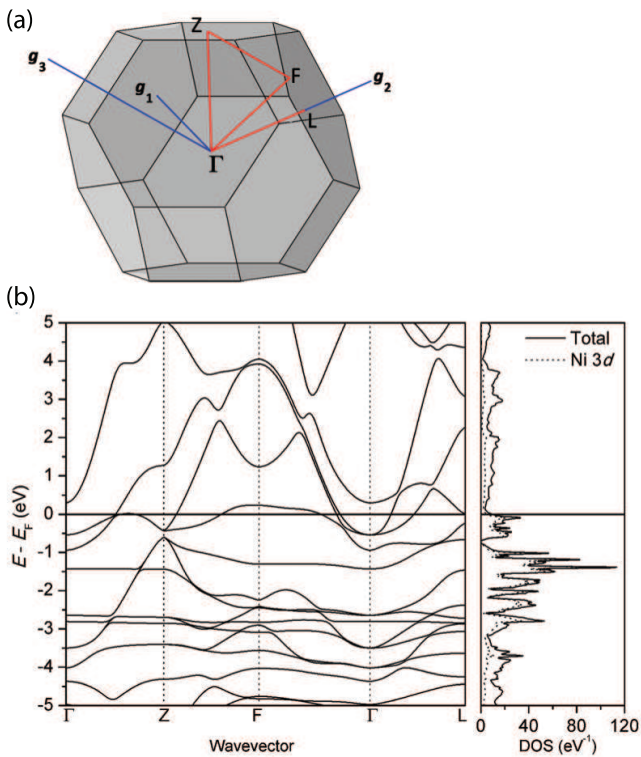}
\caption{a) First Brillouin zone of Ni$_3$Sn$_2$S$_2$ (rhombohedral setting), showing the high-symmetry k-points $\Gamma$ (0,0,0), Z (0.5,0.5,0.5), F (0,0.5,0.5) and L (0,0.5,0) used to plot the band structure. b) Calculated band-structure along high-symmetry paths and the corresponding density of states (both total and projected on the Ni 3\textit{d} states).}
\end{figure}

\begin{table}[b]
\caption{\label{tab:table1}%
Bader charges for Ni$_3$Sn$_2$S$_2$ and two reference compounds (SnS and SnSb).
}
\begin{ruledtabular}
\begin{tabular}{lcccc}
\textrm{Atom}&
\textrm{Ni}&
\multicolumn{1}{c}{\textrm{Sn}}&
\multicolumn{1}{c}{\textrm{S}}&
\multicolumn{1}{c}{\textrm{Sb}}\\
\colrule
Ni$_3$Sn$_2$S$_2$ & +0.05 & +0.64 & -0.71 & - \\
SnS & - & +0.96 & -0.96& -\\
SnSb & - & +0.36 & - & -0.36\\
\end{tabular}
\end{ruledtabular}
\end{table}
 
\subsubsection{Seebeck coefficient}
Within the constant relaxation time approximation ($\tau(\bf{k})=\tau$) in Boltzmann's transport theory, the Seebeck coefficient can be fully predicted from the DFT band structure, without introducing any empirical parameters.  Therefore the Seebeck coefficient constitutes a good test to the quality of the theoretical model. 

The calculations provide the components of the Seebeck coefficient tensor, but we find very little anisotropy (e.g. $S_{zz}/S_{xx}$ = 0.992 at 300 K). Since the layered crystal structure of shandites is clearly anisotropic, this result might seem a bit surprising, but we note that nearly isotropic Seebeck coefficients have been reported for other anisotropic crystals including Bi$_2$Te$_3$ \cite{manzano_anisotropy} and SnSe$_2$ \cite{sun_anisotropy}, at specific ranges of temperatures and doping levels. This behaviour probably results from cancellations of different contributions to the anisotropy. It is worth noting here that in our calculations we have assumed that the relaxation time is fully isotropic. However, in some cases, the differences in scattering rates in different directions may be an additional source of anisotropy \cite{fernandes_anisotropy}, which we have ignored.  Since there are no experimental data on single crystals to confirm the presence/absence of anisotropy, we will focus here on the calculated spherical average of the Seebeck coefficient, which can be compared to the experimental measurements in the polycrystalline material. 

The comparison between experimental and theoretical results is shown in Fig. 3. Two sets of experimental results are reported, which were obtained using two different instruments (one for measurements below and the other for measurements above room temperature). The discontinuity at room temperature arises from the use of different instruments, and not from physical effects. For the whole range of temperatures there is excellent agreement between theory and experiment.

\begin{figure}[!ht]
\centering
\includegraphics[width=90mm]{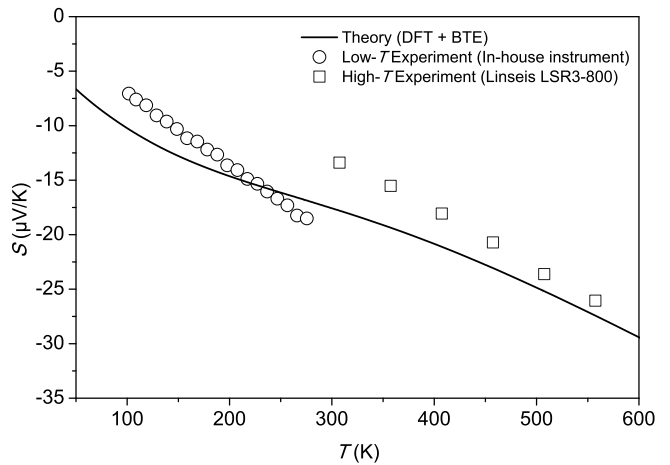}
\caption{Experimental and theoretical Seebeck coefficients as functions of temperature.}
\end{figure}

Analogous to the Co shandite, Ni$_3$Sn$_2$S$_2$ exhibits a negative Seebeck coefficient with the absolute value increasing almost linearly with temperature. However, at room temperature, the Seebeck coefficient of the Ni shandite is 3 to 4 times smaller than that of the Co analogue (ca. -50 $\mu$V/K  \cite{corps_interplay}). This can be explained by the difference in the electronic structure of the two compounds. It is well known that semiconductors generally exhibit much higher Seebeck coefficients than metals \cite{wood88}. For half-metals, the Seebeck coefficient is approximately given by the two-current model \cite{xiang, gravier_spin}, i.e. the conductivity-weighted average of the Seebeck coefficients of the two spin channels. Therefore, half-metallic Co$_3$Sn$_2$S$_2$ can be expected to have a higher Seebeck coefficient than fully metallic  Ni$_3$Sn$_2$S$_2$, as we have observed.  

 \subsubsection{Electronic conductivity and scattering rates}
In contrast to the Seebeck coefficient, the electronic conductivity can only be predicted per unit of relaxation time, i.e. at this level of theory we can only calculate the ratio $\sigma/\tau$.  There have been some recent methodological developments for the calculation of electron-phonon relaxation times from first principles \cite{Xiao2014, epw}, but the algorithms are not very mature yet and quite computationally demanding.  We have therefore chosen to combine our calculations with experimental measurements of electronic conductivity, in order to analyze the behavior of the effective isotropic relaxation time as a function of temperature.

\begin{figure}[!ht]
\centering
\includegraphics[width=90mm]{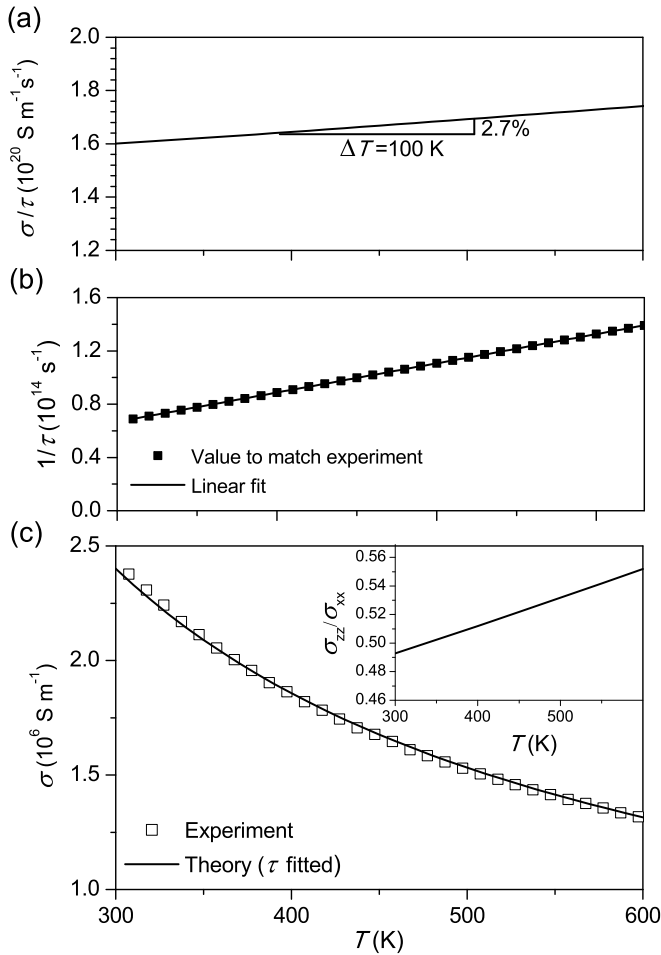}
\caption{a) Calculated electronic conductivity per unit of relaxation time ($\sigma$/$\tau$); b) electron-phonon scattering rates obtained using the experimentally determined $\sigma$ and the theoretically obtained $\sigma$/$\tau$, and linear fitting of its temperature dependence; c) experimental electronic conductivity data and calculated values using fitted $\tau(T)$. The inset shows the ratio between the \textit{zz} and \textit{xx} components of the conductivity tensor.}
\end{figure}

The $\sigma/\tau$ ratio has only a weak temperature dependence (Fig. 4a). For example, increasing the temperature from 400 to 500 K leads to an increase of less than 3\% in the value of $\sigma/\tau$.  In Fig. 4b we show the scattering rates ($1/\tau$) required to exactly match the experimental conductivities as a function of temperature (Fig. 4c).  The experimental value of $\sigma$ at room temperature is 2.4$\times$10$^{6}$ Sm$^{-1}$ and decreases with temperature as expected for a metallic system. 

The scattering rates determined in this way increase linearly with temperature. This result can be interpreted in terms of  Matthiessen's rule, according to which the total scattering rate is the sum of contributions from electron-electron scattering (proportional to $T^2$), from electron-phonon scattering (proportional to $T$ above the Debye temperature of the material), and from impurity scattering (approximately independent of $T$) \cite{singleton}. In our case, given the linearity of the dependence it is clear that the electron-electron term can be omitted and the variation can be well fitted with the linear equation:

\begin{equation}
\frac{1}{\tau}  = a_0 + a_1T
\end{equation}
for which we obtain $a_0$=9.82$\times$10$^{11}$ s$^{-1}$ and $a_{1}$=2.19$\times$10$^{11}$ s$^{-1}$K$^{-1}$. The electron-phonon term dominates at the temperatures of interest here. For example, at 300 K the impurity contribution represents less than 1.5\% of the total scattering rate, and this reduces to 0.7\% at 600K. The calculated relaxation time of 1.5$\times$10$^{-14}$ s at 300 K is reasonable and of the same order as values obtained by the same procedure in other materials (e.g. for Bi$_2$Te$_3$ \cite{Hinsche_bi2te3}). 

The linear dependence of the electron-phonon scattering rate with temperature is as expected for temperatures of the order of and above the Debye temperature of the material \cite{singleton}. In Section III.9 we provide an estimation of the Debye temperature of Ni$_3$Sn$_2$S$_2$ based on phonon calculations, and we obtain $T_{\text{D}}$=278 K, which is consistent with the present analysis. For temperatures above  $T_{\text{D}}$ and in the absence of significant impurity contributions, both $\sigma$ and $\tau$ are roughly inversely proportional to temperature, which makes the ratio $\sigma/\tau$ almost constant, as seen in Fig. 4a.

Finally we note that our calculations also provide access to individual components of the electronic conductivity tensor. On the assumption of isotropic relaxation time, we can obtain the ratio $\sigma_{zz}/\sigma_{xx}$ as a function of temperature. The inset of Fig. 4c shows that there is significant anisotropy in this case, with the conductivity within the Kagome plane ($\sigma_{xx}$) being around twice the conductivity in the perpendicular direction ($\sigma_{zz}$). This is expected since the Kagome plane contains a 2D network of zero-valent metal (Ni) atoms which locally increases the density of electronic states.

\subsubsection{Electronic thermal conductivity}
We discuss here only the theoretical calculation of the electronic contribution to the thermal conductivity ($\kappa_{\text{el}}$) (and its connection to the electronic conductivity $\sigma$), because in experiment only the total (electronic + lattice) thermal conductivity is measured. The discussion of the experimental total thermal conductivity will be presented below (Section III.9), together with results of the lattice thermal conductivity calculations.

The $\kappa_{\text{el}}$ values can also be obtained from the Boltzmann transport equation, but as in the case of $\sigma$, only the values relative to the relaxation time (i.e. $\kappa_{\text{el}}/\tau$) can be determined. The relaxation time for electronic heat transport is not necessarily the same as the relaxation time considered above for the electronic conductivity, but for metals in the regime of temperatures of interest here (similar to or above the Debye temperature) the two relaxation times can be considered approximately equal \cite{singleton}. In this case, it is expected that the ratio $\sigma/\tau$ follows the Wiedeman-Franz law, i.e. it is simply proportional to temperature, with the proportionality constant being the Lorenz number ($L_0$ =  2.44$\times$10$^{-8}$ W$\Omega$K$^{-2}$). Fig. 5 shows that the calculated transport coefficients follow the Wiedeman-Franz law to a good approximation, although there are some deviations at higher temperatures, where the effective Lorenz number becomes somewhat higher than $L_0$ (by up to 10\% at 600 K). 

The predicted absolute value for the electronic thermal conductivity (using the relaxation time determined from the experimental $\sigma$) does not vary strongly with temperature. This is expected from the Wiedemann-Franz law and the result in the previous section showing that $\sigma$ is roughly inversely proportional to temperature. We obtain $\kappa_{\text{el}}$ values of 20.8 Wm$^{-1}$K$^{-1}$ at 300 K and 22.3 Wm$^{-1}$K$^{-1}$ at 600 K. The anisotropy of the electronic thermal conductivity tensor was found to follow a very similar pattern as that of the electronic conductivity tensor. 

\begin{figure}[!ht]
\centering
\includegraphics[width=90mm]{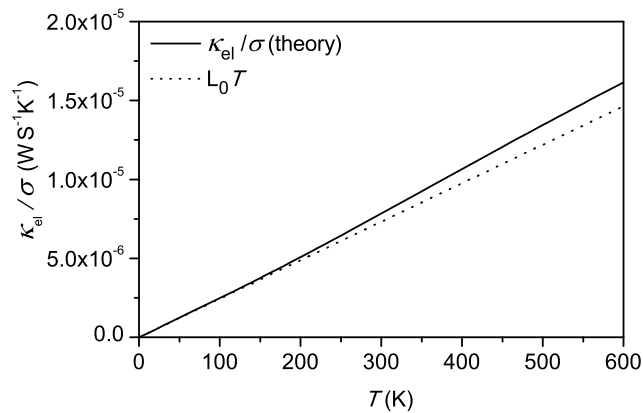}
\caption{Ratio between the electronic contribution to the thermal conductivity ($\kappa_{\text{el}}$) and the electrical conductivity ($\sigma$) as a function of temperature, in comparison with the expectation from the Wiedemann-Franz law.} 
\end{figure}

\subsubsection{Phonon structure and lattice thermal conductivity}
We now discuss the phonon behavior in Ni$_3$Sn$_2$S$_2$, as a starting point for the discussion of the lattice contribution to the thermal conductivity, 
but also to provide an estimation of the Debye temperature of the material. The phonon dispersion curves along the high-symmetry directions in the Brillouin zone are shown in Fig. 6. Consistent with the primitive cell of 7 atoms there are 21 phonon modes: 3 acoustic and 18 optical branches. The vibrational density of states is divided in two groups, the lower one comprises 15 branches and the upper one comprises 6 branches, with a gap of of about 10 meV between the two groups.

\begin{figure}[!ht]
\centering
\includegraphics[width=90mm]{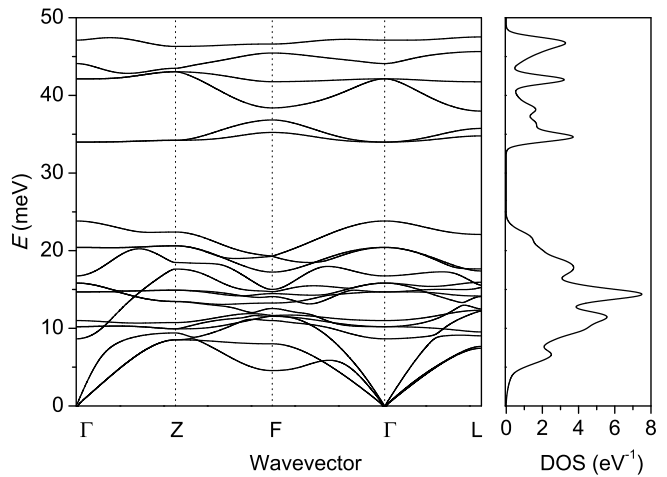}
\caption{Phonon dispersion curves of Ni$_3$Sn$_2$S$_2$ along high symmetry paths in the Brillouin zone, and the corresponding phonon density of states.}
\end{figure}

From the phonon structure we can extract the specific heat capacity of the solid as a function of temperature (ignoring for the moment anharmonic contributions), which is shown in Fig. 7. In the low-temperature limit, $C_{\text{v}}$ is proportional to  $T^3$ \cite{kittel}: 

\begin{equation}
C_\text{v}  \approx \frac{12 \pi^4}{5}Nk_{\text{B}}\left(\frac{T}{T_{\text{D}}}\right)^3
\end{equation}

from where we can calculate the Debye temperature $T_{\text{D}}$=278 K (inset of Fig. 7). 

\begin{figure}[!ht]
\centering
\includegraphics[width=90mm]{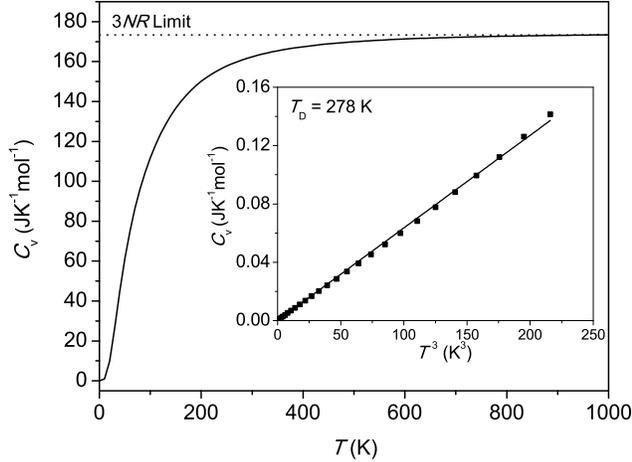}
\caption{\label{fig:epsart}Specific heat capacity of Ni$_3$Sn$_2$S$_2$ per mole of formula units. The limit value at high temperature is 3$NR$ where $N=7$ is the number of atoms per formula unit and $R$ is the gas constant. The fitting of the low- temperature values of the heat capacity to a Debye model (inset) is used to obtain the Debye temperature.}
\end{figure}

From the anharmonic displacements we can then calculate the lattice thermal conductivity $\kappa_{\text{latt}}$, which is shown in Fig. 8a as a function of temperature. At 300 K it is ca. 2 Wm$^{-1}$K$^{-1}$ and it slowly decreases with temperature down to 1 Wm$^{-1}$K$^{-1}$ at 600 K. These values are very low, well below typical values for crystalline solids, and similar to what is found for disordered materials like amorphous silicon dioxide \cite{goodson}. The origin of the low $\kappa_{\text{latt}}$ is the very anharmonic nature of the vibrations in Ni$_3$Sn$_2$S$_2$. This is reflected in a high  Gr$\ddot{\text{u}}$neisen parameter obtained from our calculations,  $\gamma$=1.55 (in the high-temperature limit), which is about three times that of Si \cite{balandin98}.

Our lattice thermal conductivity calculations provide further useful information, including directional (tensorial) components as well as the contributions from different phonon mean free paths. The $\kappa_{\text{latt}}$ tensor showed negligible anisotropy at all temperatures above 300 K. On the other hand, from the mean-free-path analysis we find that nanostructuring is not a viable strategy for reducing thermal conductivity as this would require particle sizes below $\sim$20 nm to achieve any significant effect (Fig. 8b). In any case, from a point of view of thermoelectric applications, it does not make sense to focus on reducing the lattice thermal conductivity, because the most important contribution to heat transport comes from electrons. 

The total calculated thermal conductivity, as well as the electronic and lattice contributions, are shown in Fig. 8a, in comparison with experimental measurements (only for the total thermal conductivity). $\kappa_{\text{latt}}$ contributes just $\sim$10\% of the total thermal conductivity, with the remaining 90\% resulting from electronic transport. The experimental total thermal conductivities are in excellent agreement with the theoretical values. However, it should be noted that the calculation of the electronic contribution $\kappa_{\text{el}}$  involved the use of a fitted relaxation time curve $\tau(T)$ to reproduce the electronic conductivity $\sigma(T)$. Because of the Wiedemann-Franz law, such fitting also guarantees a good theoretical value for $\kappa_{\text{el}}$. However, it is still remarkable that the calculated values of $\kappa_{\text{latt}}$, which were obtained without any fitting parameters, bring the total theoretical values of thermal conductivity to perfect agreement with experiment. Thus, our theoretical lattice thermal conductivity predictions are confirmed by the experimental measurements. 

\begin{figure}[!ht]
\centering
\includegraphics[width=90mm]{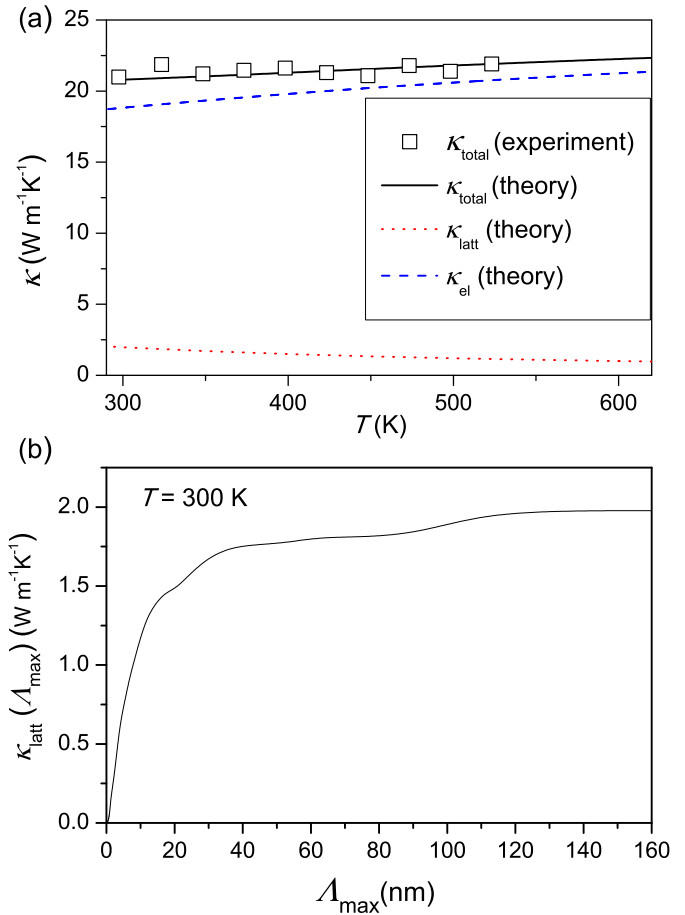}
\caption{a) Total thermal conductivities from experiment and theory, and calculated electronic ($\kappa_{\text{el}}$) and lattice ($\kappa_{\text{latt}}$) contributions versus temperature; b) $\kappa_{\text{latt}}$ at 300 K as a function of the maximum mean free path.}
\end{figure}

 \subsubsection{Thermoelectic figure of merit $ZT$}
In order to summarize the thermoelectric behavior of Ni$_3$Sn$_2$S$_2$, we have calculated the thermoelectric figure of merit $ZT$ of the material as a a function of temperature, using both theoretical and experimental data (Fig. 9a). Two theoretical $ZT$ curves are given, one excluding and the other including the lattice thermal conductivity term. In the former case, the prediction is fully ab initio, because the relaxation time cancels out, and Eq. (1) becomes:

\begin{equation}
    ZT\approx\ \frac{\sigma S^2T}{\kappa_{\text{el}}} \approx \frac{S^2}{L_0}
\end{equation}

In fact, since $\kappa_{\text{latt}}$ contributes only around 10\% to the total thermal conductivity, its effect on $ZT$ is very small, as can be seen in Fig. 9a. The calculation of $ZT$ taking into account the lattice contribution to $\kappa$ is not fully predictive, as it involved fitting of the relaxation times. In any case, the theoretical prediction agrees well with experiment, showing an increase with temperature that is approximately quadratic (because $S$ has an approximately linear variation with temperature). 

\begin{figure}[!ht]
\centering
\includegraphics[width=92mm]{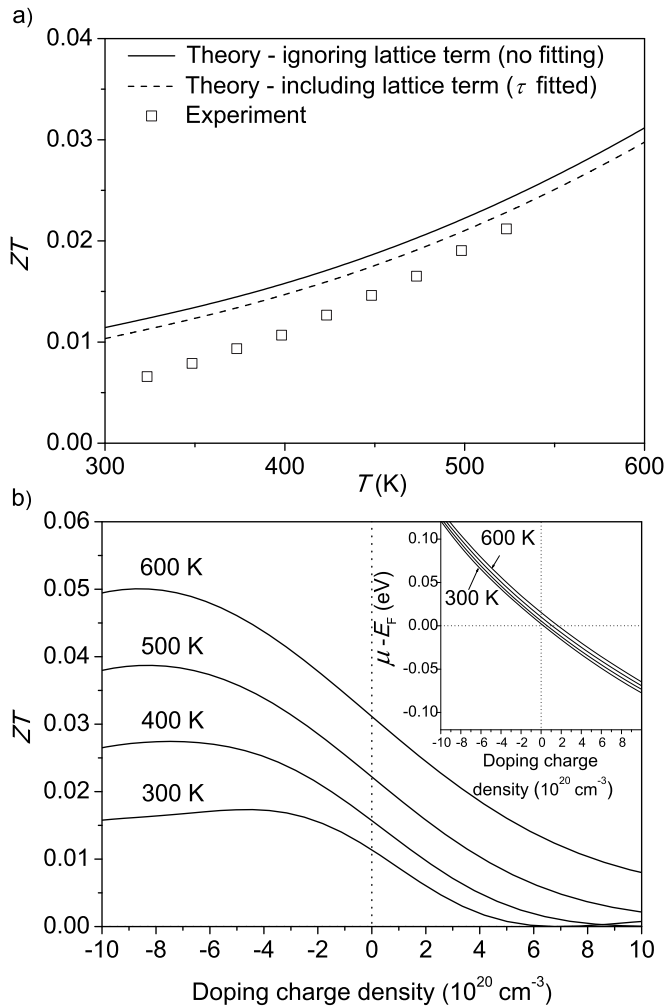}
\caption{a) Thermoelectric figure of merit ($ZT$) of Ni$_3$Sn$_2$S$_2$ from experimental and from theoretical data. b) Effect of dilute doping on $ZT$ (the inset shows the correspondence between the doping charge density and the chemical potential within the rigid band approximation).}
\end{figure}

The values of $ZT$ in Ni$_3$Sn$_2$S$_2$ are too small for thermoelectric applications, which, as Eq. 4 indicates, is mainly due to the low Seebeck coefficient. A possible strategy to improve the Seebeck coefficient and $ZT$ is via doping. We therefore consider here the response of $ZT$ to dilute doping, with either excess or deficiency of electrons, within a rigid band approach. Fig. 9b shows the dependence of $ZT$ on the concentration of doped charge carriers, which can be studied by varying the chemical potential, as shown in the figure inset. $ZT$ is predicted to increase with small concentrations of electron doping (negative doping charge density), although the magnitude of the increase is still modest. Clearly, a wider range of doping concentrations must be considered in trying to achieve a significant increase in $ZT$. However, the theoretical investigation of heavily doped Ni$_3$Sn$_2$S$_2$ is beyond the scope of the present study, as it cannot be rigorously done within the rigid band approach; it would instead require to include the dopant atoms explicitly in the supercell (e.g. \cite{suzuki15, CuCoO2}).

\section{Conclusions}
We have reported a full theoretical and experimental study of  the electronic, phonon structure and transport behavior of the shandite-type Ni$_3$Sn$_2$S$_2$. The theoretical results, in particular the Seebeck coefficient and lattice thermal conductivities, which are predicted without any fitting parameters, are in excellent agreement with experiment. The dominant electron scattering mechanism is via phonons, and from the comparison of theoretical and experimental results we have obtained the temperature dependence of the electron scattering rate. 

Pure Ni$_3$Sn$_2$S$_2$ is not a good thermoelectric material, as it has a very low thermoelectric figure of merit ($ZT\sim10^{-2}$ at room temperature). Its lattice thermal conductivity is very low, and contributes only $\sim10\%$ of the total thermal conductivity. Therefore, for this material very little can be gained by nanostructuring or other strategies aimed at reducing heat transport by phonons. In fact, for this metallic shandite, $ZT$ is mainly a function of the Seebeck coefficient. In order to improve $ZT$, a dramatic change in the Seebeck coefficient would be needed, which we show it cannot be achieved by dilute doping. The effort in finding thermoelectric shandites should clearly focus on the half-metallic or semiconductor systems, where the Seebeck coefficients can be engineered to much higher values.  

\section{\label{sec:level1}Acknowledgements\protect\\}

AA and PM acknowledge funding from EPSRC DTG studentships. Calculations were performed using the UK National Supercomputing Facility ARCHER via our membership of the U.K.’s HPC Materials Chemistry Consortium (EPSRC Grant EP/L000202). The authors wish to thank the University of Reading for access to the Chemical Analysis Facility for powder X-ray diffraction measurements. 




\bibliography{Ni3Sn2S2}
\bibliographystyle{apsrev4-1}

\end{document}